\documentclass[aps,amssymb,article,pra,twocolumn,showpacs,showkeys,floatfix]{revtex4}

\usepackage{epsf}
\usepackage{amsmath,bm}
\usepackage{graphicx}
  \begin{document}
         
  \title{Anomalous particles}

  \author{Boris I. Ivlev}

  \affiliation{Instituto de F\'{\i}sica, Universidad Aut\'onoma de San Luis Potos\'{\i},\\ 
  San Luis Potos\'{\i}, 78000 Mexico}

  \begin{abstract}
  In the whole set of solutions of the Dirac equation there is a distinct class referred to as anomalous. Corresponding anomalous particles are independent of conventional ones. 
  The concept of anomalous particles is applicable to Dirac insulators, where electrons obey the Dirac like wave equation. Positively charged antielectrons, which are not 
  holes, can exist in the Dirac insulator. In this material one can create the electron-antielectron pair keeping the valence band completely filled. The anomalous subsystem, 
  associated with the electron-antielectron vacuum, is an inner property of the Dirac insulator. Anomalous quasiparticles in the Dirac insulator can be identified in 
  experiments with electric current. 
  
  \end{abstract} \vskip 1.0cm



  \maketitle

  \section{INTRODUCTION}
  \label{intr}
  The relativistic quantum mechanics of electrons and positrons is described by the Dirac equation \cite{LANDAU2}. This equation is also applicable in the field of Dirac 
  materials in condensed matter \cite{MAT}. Solutions of the Dirac equation are well studied \cite{LANDAU2,DIM,PAL}. However, it is revealed that in the whole set of 
  solutions of the Dirac equation there is additionally a distinct class referred to as anomalous. Features of anomalous and conventional states are different.
  
  For a free particle the Dirac plane wave is a particular superposition of angular harmonics $\sum c_n\psi_n(\pmb r)$ with the quantum number $n$ consisted of the energy 
  $\varepsilon_q$, the total angular momentum $j$, its projection $m$, and the orbital momentum $l$ ($\psi_n\sim r^l$ on short distance) \cite{LANDAU2}. It happens that there 
  exists another class of states, where the expansion coefficients $c_n$ are coupled differently, $c_n\sim (1/r_c)^l$. Here $r_c$ is the length parameter. This expansion in 
  spherical eigenfunctions is convergent solely at $r<r_c$. Outside this region the certain generating function of coordinates provides the analytical continuation to $r>r_c$. 
  The continued wave function is of the form $\sum \tilde c_n\tilde\psi_n(\pmb r)$, where $\tilde\psi_n\sim 1/r^{l+1}$ on short distance. This crossover occurs not for each 
  $\psi_n(\pmb r)$ but in terms of $l$-powers of coordinates formed by sets of $m$. 
  
  That anomalous state is not a superposition of conventional eigenstates contrary to a usual case. For a free anomalous particle the formal expansion of its wave function in 
  Fourier plane waves contains harmonics with $|\pmb p|\neq q$, which are not conventional eigenstates. 
  
  The finite radius of convergence $r_c$ makes the anomalous state irreducible to a conventional one. Despite the electron may be of a non-relativistic energy, one should 
  apply the Dirac formalism. The different mathematical aspect is that some stationary solutions of the Hamiltonian wave equation cannot be expanded in conventional eigenfunctions.
  
  The Dirac insulator is a special type of matter, where electrons obey the Dirac like wave equation \cite{PAN,KAN,MAT,WOL,FUS}. The topological aspect of Dirac insulators is 
  not a whole story. In those materials the anomalous subsystem is formed allowing anomalous electrons and antielectrons. The antielectron is a positively charged quasiparticle, 
  which is not a vacancy in the valence band. Like the positron is not a vacancy in ``Dirac sea'' but can be created from the electron-positron vacuum. Analogously, the Dirac 
  insulator associates with the electron-antielectron vacuum. One can create the electron-antielectron pair keeping the valence band completely filled. The known transition 
  from conventional insulator to topological one \cite{XUI} is assisted by the formation of the anomalous subsystem. This is an inner property of bulk Dirac insulators. 
  
  Anomalous quasiparticles in the Dirac insulator can be identified in experiments with electric current across a border between the Dirac and conventional insulators. 
  
  \section{CONVENTIONAL ELECTRON STATES}
  \label{wave}
  The wave function of a free electron obeys the Dirac equation \cite{LANDAU2}
  \begin{equation}
  \label{1} 
  \left(i\gamma^0\frac{\partial}{\partial t}+ic\pmb\gamma\cdot\nabla -mc^2\right)\psi(\pmb r,t)=0,
  \end{equation}
  where $\pmb r=(\pmb\rho,z)$. The bispinor $\psi$ and $\gamma$-matrices are
  \begin{equation}
  \label{2} 
  \psi=
  \begin{pmatrix}
  \Phi\\
  \Theta 
  \end{pmatrix},
  \hspace{0.5cm}\pmb\gamma=
  \begin{pmatrix}
  0&\pmb\sigma\\
  -\pmb\sigma&0
  \end{pmatrix},
  \hspace{0.5cm}\gamma^0=
  \begin{pmatrix}
  1&0\\
  0&-1
  \end{pmatrix}.
  \end{equation}
  In spherical coordinates the eigenstates are marked by the quantum number $n$ consisted of the energy $\varepsilon_q$, the total angular momentum $j$, its 
  projection $m$, and the orbital momentum $l$. With the quantum numbers $j=3/2$, $l=1$, and $|m|\leq j$ the solutions are
  \cite{LANDAU2}
  \begin{equation}
  \label{12}
  \Phi_n(\pmb r)=\sqrt{\frac{\varepsilon_q+mc^2}{6\varepsilon_q}}R_{q1}(r)
  \begin{pmatrix}
  \sqrt{3/2+m}\,Y_{1,m-1/2}\\
  \sqrt{3/2-m}\,Y_{1,m+1/2}
  \end{pmatrix}\\
  \end{equation}
  \begin{equation}
  \Theta_n(\pmb r)=\sqrt{\frac{\varepsilon_q-mc^2}{10\varepsilon_q}}R_{q2}(r)
  \begin{pmatrix}
  \sqrt{5/2-m}\,Y_{2,m-1/2}\\
  -\sqrt{5/2+m}\,Y_{2,m+1/2}
  \end{pmatrix}.
  \label{12a}
  \end{equation}
  The radial functions $R_{ql}(r)$ are \cite{LANDAU2}
  \begin{equation}
  \label{15}
  R_{q1}(r)=2q\left(\frac{\sin qr}{q^2r^2}-\frac{\cos qr}{qr}\right),\hspace{0.5cm} R_{q2}(r)\sim q^3r^2.
  \end{equation}
  
  A variety of solutions of the Dirac equation is described, for instance, in \cite{DIM}. It is surprising that there are the certain states, referred to as 
  anomalous, which are outside the conventional scheme that is they are not superpositions of conventional eigenstates.
  
  \subsection{Transformation of the wave function} 
  \label{tran}
  One can make the transformation  
  \begin{equation}
  \label{3} 
  \psi=\exp\left[\begin{pmatrix}
  0&\sigma_z\\
  \sigma_z&0
  \end{pmatrix}\lambda\right]\psi'=\left[\cosh\lambda + \begin{pmatrix}
  0&\sigma_z\\
  \sigma_z&0
  \end{pmatrix} \sinh\lambda\right]\psi'
  \end{equation}
  of the wave function, where $\lambda$ is a parameter. The transition from $\psi$ to $\psi'$ is given by the same formula with the formal change 
  $\lambda\rightarrow -\lambda$.
  
  Note that the Lorentz transformation of coordinates and potentials to the frame, moving with the velocity $v_z=v$, should be supplemented by the transformation 
  (\ref{3}) with 
  \begin{equation}
  \label{4}
  \lambda=\frac{1}{4}\ln\frac{c+v}{c-v}
  \end{equation}
  to get the invariant form of the Dirac equation \cite{AKH}. 
  
  Below we use the parametrization (\ref{4}) just to introduce the auxiliary function $\psi'$ instead of the physical wave function $\psi$. At $v/c\ll 1$ it follows 
  from (\ref{3}) that 
  \begin{equation}
  \label{2aa} 
  \begin{pmatrix}
  \Phi'\\
  \Theta' 
  \end{pmatrix}=
  \begin{pmatrix}
  \Phi -\sigma_z\Theta v/2c\\
  \Theta-\sigma_z\Phi v/2c 
  \end{pmatrix}.
  \end{equation}
  
  The Dirac equation (\ref{1}) for $\psi(\pmb r,t)=\psi(\pmb r)\exp(-it\varepsilon_q)$, with the transformation (\ref{3}), acquires the form
  \begin{eqnarray}
  \nonumber
  &&\Bigg[\left(\gamma^0+\frac{v}{c}\gamma_z\right)\varepsilon_q-m'c^2\\
  &&+ic\left(\pmb\gamma\cdot\nabla'+\frac{v}{c}\gamma^0\frac{\partial}{\partial z}\right)\Bigg]\psi'(\pmb r')=0,
  \label{5}
  \end{eqnarray}
  where $\pmb r'=(\pmb\rho',z)$. The definitions $\pmb\rho'=\pmb\rho/\sqrt{1-v^2/c^2}$ and $m'=m\sqrt{1-v^2/c^2}$ are introduced. 
  
  Analogously to (\ref{2}), the bispinor $\psi'$ consists of two spinors. The equations for them follow from (\ref{5})
  \begin{equation}
  \left(\varepsilon_q-m'c^2+iv\frac{\partial}{\partial z}\right)\Phi'=-\left(ic\pmb\sigma\cdot\nabla'+\frac{v}{c}\varepsilon_q\sigma_z \right)\Theta',
  \label{6}
  \end{equation}
  \begin{equation}
  \left(\varepsilon_q+m'c^2+iv\frac{\partial}{\partial z}\right)\Theta'=-\left(ic\pmb\sigma\cdot\nabla'+\frac{v}{c}\varepsilon_q\sigma_z \right)\Phi'.
  \label{7}
  \end{equation}
  
  If to introduce the spinor
  \begin{equation}
  G(\pmb\rho',z)=\left(\pmb\sigma\cdot\nabla'-\frac{iv}{c^2}\varepsilon_q\sigma_z\right)\Theta'(\pmb\rho',z)
  \label{8}
  \end{equation}
  the solution of Eq.~(\ref{6}) is
  \begin{equation}
  \Phi'_{1,2}(\pmb\rho',z)=\frac{c}{v}\int^{\infty}_{z}d\xi\exp\left(i\frac{z-\xi}{L}\right)G_{1,2}(\pmb\rho',\xi),
  \label{9}
  \end{equation}
  where
  \begin{equation}
  L=\frac{v}{\varepsilon_q-m'c^2}.
  \label{10}
  \end{equation}
  In the limit $v\rightarrow 0$ the form (\ref{9}) smoothly goes over into the conventional solution.
  
  Eq.~(\ref{7}) in spinor components has the form
  \begin{eqnarray}
  \label{11} 
  &&\left(\varepsilon+m'c^2+iv\frac{\partial}{\partial z}\right)  
  \begin{pmatrix}
  \Theta_{1}'\\
  \Theta_{2}' 
  \end{pmatrix}\\
  &&=-ic\begin{pmatrix}\left(\partial/\partial x'-i\partial/\partial y'\right)\Phi_2'+(\partial/\partial z-iv\varepsilon/c^2)\Phi'_{1}\\ \\
  \left(\partial/\partial x'+i\partial/\partial y'\right)\Phi_1'-(\partial/\partial z-iv\varepsilon/c^2)\Phi'_{2}\end{pmatrix}.
  \nonumber
  \end{eqnarray}
  Now one should insert (\ref{9}) into Eqs.~(\ref{11}) to obtain equations for the components $\Theta_{1,2}'$. 
  
  \section{$\pmb\eta$-GENERATED SOLUTIONS}
  \label{ano}
  Eq.~(\ref{6}) has a different solution besides (\ref{9}). This solution is
  \begin{equation}
  \Phi'_{1,2}(\pmb\rho',z)=\frac{c}{v}\int^{\eta_{1,2}}_{z}d\xi\exp\left(i\frac{z-\xi}{L}\right)G_{1,2}(\pmb\rho',\xi),
  \label{9a}
  \end{equation}
  where the field $\pmb\eta=\{\eta_{1}(\pmb\rho'),\eta_{2}(\pmb\rho')\}$ is real. Otherwise it could be a grow on large distance. The typical spatial scale
  $\rho'$ of $\pmb\eta$ is the certain length $r_c$. The solution (\ref{9a}) can be referred to as $\pmb\eta$-generated solution. 
  
  \subsection{Short distance $r\ll r_c$}
  \label{short} 
  Below we consider the limit $v\ll c$. At $r,\eta_{1,2}\ll L$ one can substitute the exponent in (\ref{9a}) by unity. Under these conditions, as follows from 
  (\ref{8}) and (\ref{9a}), $\Phi'\sim (c/v)\Theta'$ and (\ref{11}) is reduced to its derivative part in the right-hand side.
  
  The difference between $\pmb r'$ and $\pmb r$, proportional to $v^2/c^2$, is neglected and (\ref{11}) takes the form
  \begin{eqnarray}
  \label{18}
  \left(\frac{\partial}{\partial x}-i\frac{\partial}{\partial y}\right)\int^{\eta_2(\pmb\rho)}_{z}d\xi G_2(\pmb\rho,\xi)=G_1(\pmb\rho,z),\\
  \left(\frac{\partial}{\partial x}+i\frac{\partial}{\partial y}\right)\int^{\eta_1(\pmb\rho)}_{z}d\xi G_1(\pmb\rho,\xi)=-G_2(\pmb\rho,z).
  \label{19}
  \end{eqnarray}
  
  Below we consider $r_c<L$. On the short distance $r<r_c$ the linear approximation  holds
  \begin{equation}
  \label{21}
  \eta_1(\pmb\rho)=ax+by,\hspace{0.5cm}\eta_2(\pmb\rho)=cx+dy.
  \end{equation}
  It follows from (\ref{18}) and (\ref{19}) that
  \begin{eqnarray}
  \nonumber
  &&G_1(0,0)-(c-id)G_2(0,0)=0,\\
  &&(a+ib)G_1(0,0)+G_2(0,0)=0.
  \label{22}
  \end{eqnarray}
  The condition of consistency of (\ref{22}) is $(a+ib)(c-id)+1=0$. The additional condition of reality of the coefficients results in the relations for the generating
  functions
  \begin{equation}
  \label{23}
  \eta_1(\pmb\rho)=ax+by,\hspace{0.4cm}\eta_2(\pmb\rho)=-\frac{ax+by}{a^{2}+b^{2}},\hspace{0.4cm}\rho\ll r_c.
  \end{equation}
 
  As follows from (\ref{22}),
  \begin{equation}
  \label{24}
  G(0,0)=\tilde C\frac{v}{c}\begin{pmatrix}
  a-ib\\
  -a^2-b^2 
  \end{pmatrix},
  \end{equation}
  where $\tilde C$ is a constant. 
  
  One obtains from (\ref{9a}) at the linear approximation on $r$
  \begin{eqnarray}
  \nonumber
  &&\Phi'_1=\tilde C(a-ib)(ax+by-z)\\
  &&\Phi'_2=\tilde C\left[ax+by+z(a^2+b^2)\right].
  \label{24a}
  \end{eqnarray}
  As follows from (\ref{24}) and (\ref{8}), at small $r$
  \begin{equation}
  \label{13}
  \Theta'=\tilde C(a-ib)\frac{v}{c}
  \begin{pmatrix}
  \alpha_1x+\beta_1y+(1-\alpha_2+i\beta_2)z\\
  \alpha_2x+\beta_2y+(a+ib+\alpha_1+i\beta_1)z
  \end{pmatrix},
  \end{equation}
  where $\alpha_{1,2}$ and $\beta_{1,2}$ are constants. According to (\ref{2aa}), $\Phi=\Phi'$ if to omit $(v/c)^2$ corrections. 
  
  Let us introduce the solution of the Dirac equation (\ref{1})
  \begin{equation}
  \label{24d}
  \psi_g(\pmb r)=C\sum^{3/2}_{m=-3/2}A_m\psi_{q,3/2,1,m}(\pmb r).
  \end{equation}
  With the relations $rY_{1,0}=iz\sqrt{3/4\pi}$ and $rY_{1,\pm 1}=\mp i(x\pm iy)\sqrt{3/8\pi}$ \cite{LANDAU1}, on short distance the upper spinor in (\ref{24d}) is
  \begin{eqnarray}
  \nonumber
  &&\frac{\Phi_g(\pmb r)}{\tilde C}= A_{-3/2}\begin{pmatrix}
  0\\
  \sqrt{3}(x-iy)
  \end{pmatrix}
  +A_{-1/2}\begin{pmatrix}
  x-iy\\
  2z
  \end{pmatrix}\\
  &&+A_{1/2}\begin{pmatrix}
  2z\\
  -(x+iy)
  \end{pmatrix}
  +A_{3/2}\begin{pmatrix}
  -\sqrt{3}(x+iy)\\
  0
  \end{pmatrix}.
  \label{13b}
  \end{eqnarray}
  According to (\ref{12a}), on short distance $\Theta_g\sim q^3r^2$.
  
  With the connection of constants
  \begin{equation}
  \label{24e}
  \tilde C=C\frac{iq^2}{6}\sqrt{\frac{\varepsilon_q+mc^2}{\pi\varepsilon_q}}
  \end{equation}
  the $\pmb\eta$-generated solution (\ref{24a}) matches the solution of the Dirac equation (\ref{24d}) on short distance under the conditions 
  \begin{eqnarray}
  \nonumber
  &&A_{-3/2}=\frac{a+ib}{2\sqrt{3}},\hspace{0.4cm}A_{-1/2}=\frac{a^2+b^2}{2},\\
  &&A_{1/2}=-\frac{a-ib}{2},\hspace{0.4cm}A_{3/2}=-\frac{(a-ib)^2}{2\sqrt{3}}.
  \label{14b}
  \end{eqnarray}
  
  In the linear approximation on $r$, $\Theta_g=0$ or $\Theta'=-\sigma_z\Phi'v/2c$ as follows from (\ref{2aa}). This condition holds if to put 
  \begin{equation}
  \label{13a}
  \alpha_1=-\frac{a}{2},\hspace{0.3cm}\beta_1=-\frac{b}{2},\hspace{0.3cm}\alpha_2=\frac{a}{2(a-ib)},\hspace{0.3cm}\beta_2=\frac{b}{a}\,\alpha_2.
  \end{equation}
  
  The solution of the Dirac equation (\ref{24d}), generally, depends on eight independent real parameters. The $\pmb\eta$-solution (\ref{13b}) depends on 
  four independent real parameters (the complex coefficient $C$ provides two of them). 
  
  In the state (\ref{24d}) the projection of total angular momentum is zero because 
  \begin{equation}
  \label{14bb}
  -\frac{3}{2}|A_{-3/2}|^2-\frac{1}{2}|A_{-1/2}|^2+\frac{1}{2}|A_{1/2}|^2+\frac{3}{2}|A_{3/2}|^2=0,
  \end{equation}
  as follows from (\ref{14b}). This condition is reduced to the compensation of $-3/2,1/2$ (or $-1/2,3/2$) components. 
  
  \subsection{Arbitrary distance $r\sim r_c$}
  \label{arb}
  If to express the obtained $\pmb\eta$-generated solution $\psi'$ through $\psi$ according to (\ref{3}), that wave function $\psi$ should obey the Dirac equation 
  (\ref{1}). But the Dirac wave function (\ref{24d}) satisfies Eq.~(\ref{9a}) solely in the lowest order on coordinates. In order to match in all orders one should 
  supplement (\ref{24d}) by higher order eigenfunctions
  \begin{equation}
  \label{23bc}
  \psi(\pmb r)=\sum_nc_n\psi_{n}(\pmb r),
  \end{equation}
  where the quantum number $n=(\varepsilon_q,j,l,m)$ and the summation occurs on $l\geq 1$ and $|m|\leq j=l+1/2$. The term with $l=1$ coincides with (\ref{24d}). 
  
  For each chosen generating field  $\pmb\eta=\{\eta_1(x,y),\eta_2(x,y)\}$ (with the dependence (\ref{23}) on short distance) one can determine the wave function. 
  The expansion of $\eta_{1,2}(x,y)$ in powers of coordinates is
  \begin{eqnarray}
  \nonumber
  &&\eta_{1}(x,y)=\sum^{\infty}_{l=1}\sum^{l}_{p=0}\alpha_{1}(l,p)x^{l-p}y^p,\\
  &&\eta_2(x,y)=\sum^{\infty}_{l=1}\sum^{l}_{p=0}\beta_l(p)x^{l-p}y^p,
  \label{9dda}
  \end{eqnarray}
  where $\alpha_1(0)=a$, $\alpha_1(1)=b$, $\beta_1(0)=c$, and $\beta_1(1)=d$. The contribution to the generating function $\eta_{1,2}(x,y)$ on the order of $l$ contains 
  $2l+2$ constants $\alpha_l(p)$ and $\beta_l(p)$ resulting in the $l$-th order of $\Phi'(x,y,z)$ (and thus of $\Phi$), which depends on $2j+1=2l+2$ independent constants 
  (the condition $\pmb\sigma\cdot\nabla\Phi^{(2)}=0$ reduces the general number of polynomial coefficients in $\Phi^{(l)}(x,y,z)$ to $2l+2$). 
  
  For example, at $l=2$ the spinor $\Phi'^{(2)}$ contains six independent constants. The spinor $G^{(1)}$ containd six constants, $\eta_1$ together with $\eta_2$ contain 
  six constants. Equalizing the terms $z$ and $z^2$, we express the six constants of $G^{(1)}$ through the six constants of $\Phi'^{(2)}$. Then, equalizing $x^2$, $y^2$, 
  and $xy$ terms, the six constants of $\Phi'^{(2)}$ are expressed through the six constants of $\eta_{1,2}$. 
  
  Such procedure can be done for each order $l$. This way, the spinor $\Phi(x,y,z)$ is expressed through the given functions $\eta_1(x,y)$ and $\eta_2(x,y)$. 
  
  The functions $\eta_{1,2}(x,y)$ are expanded in powers of $x/r_c$ and $y/r_c$, where $r_c\ll L$ is the typical length scale. At $r\sim r_c$ one can substitute the 
  exponent in (\ref{9a}) by unity
  \begin{equation}
  \Phi'_{1,2}(\pmb\rho,z)=\frac{c}{v}\int^{\eta_{1,2}}_{z}d\xi \,G_{1,2}(\pmb\rho,\xi)
  \label{9as}
  \end{equation}
  with the definition $G=\pmb\sigma\cdot\nabla\Theta'$.
  
  At $v\ll c$, $\Phi'\sim (c/v)\Theta'$. Under this condition the term $\pmb\sigma\cdot\nabla\Phi$ in (\ref{7}) dominates due the large coefficient $c/v$ and should be 
  equalized to zero. Thus, Eqs.~(\ref{18}) and (\ref{19}) are valid in our case of a non-linear $\eta(x,y)$. Its expansion in powers determines the expansion of the wave 
  function. According to the relation $\Phi=\Phi'+\sigma_z\Theta' v/2c$, $\Phi=\Phi'$ if to drop the $(v/c)^2$ correction. 
  
  \section{ANOMALOUS PARTICLES}
  \label{anst}
  Now one can study how a choice of the generating field determines properties of resulting particles. 
  
  If to choose the generating field in the form
  \begin{equation}
  \pmb\eta(x,y)=(ax+by)\exp\left(\frac{x^2+y^2}{r^{2}_{c}}\right)\left\{1,-\frac{1}{a^2+b^2}\right\}
  \label{11da}
  \end{equation}
  the powers of the wave function go with the coefficients $\sim 1/n!$. This provides convergence of the series for $\Phi'$ at all coordinates. Hence, the function 
  $\Phi'(x,y,z)$ (and thus $\Phi(x,y,z)$) can be expanded in powers of distance in the entire space. The wave function, formed this way, is just a superposition of 
  angular harmonics. 
  
  The situation becomes qualitatively different if to choose the generating field in the pole form 
  \begin{equation}
  \pmb\eta(x,y)=\frac{ax+by}{1+(x^2+y^2)/r^{2}_{c}}\left\{1,-\frac{1}{a^2+b^2}\right\}.
  \label{10da}
  \end{equation}
  Suppose that $y=z=0$ and $x$ is continued to the complex plane, where it is close to $ir_c$. Then $\eta_{1,2}\sim 1/(x-ir_c)$ becomes large close to the pole. 
  Suppose now that the wave function is a regular function of coordinates that is it can be expanded in powers of coordinates in the entire space. In this case, the 
  integral in (\ref{9a}) should be also regular functions of $(x-ir_c)$. But this is impossible since it is of the singular form 
  \begin{equation}
  \exp\left(-\frac{ia}{2L}\frac{r^{2}_{c}}{x-ir_c}\right).
  \label{10dh}
  \end{equation}
  Therefore the wave function is not a regular function and cannot be expanded in powers of coordinates in the entire space. Also the wave function cannot be singular 
  at real coordinates since each function $\int G_{1,2}(x,y,\xi)d\xi$ in (\ref{18}) and (\ref{19}) satisfies the three-dimensional Laplace equation. 
  
  It follows that a singularity of the generating field $\pmb\eta(x,y)$ at complex coordinates results in the assisting singularity of the wave function. Its expansion 
  in powers of coordinates is convergent solely at the distance less than the certain radius of convergence $r_{c}$. That is the wave function is a superposition of 
  eigenfunctions (\ref{23bc}) on the short distance $r<r_c$ only. At $r>r_c$ the expansion goes in powers of $r_c/r$. The crossover can be represented as  
  \begin{equation}
  \sum_nc_n\psi_n(\pmb r)\Big|_{r<r_c}\rightarrow\sum_n\tilde c_n\tilde\psi_n(\pmb r)\Big|_{r>r_c},
  \label{12da}
  \end{equation}
  where the initial part is (\ref{23bc}). The functions $\psi\sim r^l$ and $\tilde\psi\sim 1/r^{l+1}$ on short distance. In (\ref{12da}) $n=(\varepsilon_q,j,l,m)$ and 
  the summation occurs on $l\geq 1$ and $|m|\leq j=l+1/2$. 
  
  Let us schematically demonstrate how the crossover (\ref{12da}) is formed. For radial wave functions it should be 
  \begin{equation}
  \sum_lc_lJ_{l+1/2}(qr)\Big|_{r<r_c}\rightarrow\sum_l\tilde c_lN_{l+1/2}(qr)\Big|_{r>r_c}
  \label{12dc}
  \end{equation}
  with the summation on even $l=2k$. The expression of the radial wave function $R_{ql}(r)\sim Z_{l+1/2}(qr)/\sqrt{r}$ through the Bessel function is used \cite{LANDAU1}. 
  At finite $z$ and $l\rightarrow\infty$ there are evaluations $J_{l}(z)\sim(z/l)^l$ and $N_{l}(z)\sim(l/z)^l$. The coefficients $c_l\sim(l/qr_c)^l$ relate to the expansion
  $\sum_k (-r^2/r^{2}_{c})^k$ at $r<r_c$ in (\ref{12dc}). The related coefficients $\tilde c_l\sim(qr_c/l)$ provide the matching expansion $\sum_k (-r^{2}_{c}/r^2)^k$ 
  at $r>r_c$. This large $l$ limit corresponds to the wave function $1/(r^2+r^{2}_{c})$ close to the pole $r=ir_c$ analogous to (\ref{10da}).
  
  The crossover (\ref{12da}) is moderated by the given generating field of the type (\ref{10da}). The resulting wave function is referred to as anomalous. This function 
  is not a superposition of conventional eigenfunctions in the entire space contrary to the case (\ref{11da}) of a regular $\pmb\eta(x,y)$. 
  
  One concludes that the analytical properties of the generating field $\pmb\eta(x,y)$ determine a class of solutions of the Dirac equation. It can be either a 
  superposition of conventional eigenfunctions (no singularities of $\pmb\eta$ at complex coordinates) or so called anomalous wave function (with singularities of 
  $\pmb\eta$), which is not reduced to such superposition. In other words, the anomalous subsystem parallels conventional one. 
  
  It could be various types of singularities of $\pmb\eta$ at various points in the complex plane. It is not clear what are ``quantum numbars'' and ``density of
  states'' since each given singular field $\pmb\eta(x,y)$ genertaes a particular shape of the anomalous wave function.
  
  Each $\psi_n(\pmb r)$ can be expanded in Dirac $\pmb p$ plane waves proportional to $\delta(|\pmb p|-q)$ \cite{LANDAU2}. The formal Fourier expansion of the 
  anomalous wave function contains plane waves with $|\pmb p|\neq q$. This does not contradict the Dirac equation since the expansion $\sum c_n\psi_n(\pmb r)$
  is not extended to the entire space. Thus the anomalous wave function cannot be expanded either in spherical harmonics or in Dirac plane waves. 
  
  The velocity parameter $v$ is involved into formation of the anomalous wave function. It satisfies the condition $r_c(\varepsilon_q-mc^2)<v<c$. 
  
  The anomalous electron and positron are independent of conventional ones having the continuous energy spectrum $\varepsilon>\Delta$. A conventional electron-positron 
  pair can be created by a photon from the electron-positron vacuum. Analogously an anomalous pair can be created from the anomalous vacuum. 
  
  For the usual Schr\"{o}dinger equation one can try to form the anomalous wave function as a superposition of independent spherical eigenfunctions
  \begin{equation}
  \psi(\pmb r)=\sum_{l,m}c_{l,m}R_{ql}Y_{l,m}(\theta,\varphi). 
  \label{12db}
  \end{equation}
  But without a binding by the natural source like $\pmb\eta$ each of these eigenfunctions ``knows nothing'' about other components and thus independently extends to the 
  entire space. The resulting infinite wave function is non-physical. The anomalous wave functions are absent in the Schr\"{o}dinger formalism since the generating field 
  is associated with Dirac like components. 
  
  The anomalous Majorana wave function is
  \begin{equation}
  \label{26bb}
  \psi^{M}(\pmb r)=\frac{1}{\sqrt{2}} 
  \begin{pmatrix}
  1&\sigma_y\\
  \sigma_y&-1
  \end{pmatrix}\psi(\pmb r),
  \end{equation}
  where $\psi(\pmb r)$ is the Dirac anomalous wave function. The wave function $\psi^{M}(\pmb r)$ satisfies the Majorana wave equation, where a particle coincides with 
  its antiparticle \cite{LANDAU2}. As one can see, anomalous Majorana particles are also possible. 
  
  In the massless case one can apply the Weyl transformation to $\psi(\pmb r)$ to obtain the anomalous Weyl particles.
  
  \section{ANOMALOUS PARTICLES IN DIRAC MATERIALS}
  \label{mat}
  Dirac materials is a class of condensed matter \cite{MAT}. Dirac insulators is a class of Dirac materials, when the electron in a crystal lattice does not obey the 
  Schr\"{o}dinger formalism but rather Dirac one \cite{PAN,KAN,MAT,WOL,FUS}. In these systems \cite{NIM,BERN,MAT} conduction and valance bands are formed self-consistently.
  Electron and hole are interconnected particles linked by the Dirac equation with the certain velocity $\tilde c\sim 10^{7}cm/s$ instead of speed of light. The concept 
  of Dirac insulator is extended to the finite gap, which is analogous to the double mass. 
  
  The examples of bulk topological insulator are $\rm{BiTl(S_1Se_0)_2}$ \cite{XUI} and $\rm{Na_3Bi}$ \cite{LIU}. The electron wave function $\psi_{\pmb q}$ is defined in the 
  Brillouin zone (BZ). At small momentum $q\ll 1/a$, where $a$ is size of the unit cell, the spinor wave function obeys the Dirac equation \cite{PAN,MAT,WOL,FUS}
   \begin{equation}
  \label{46}
  \left(\gamma^0\varepsilon-\tilde c\,\pmb\gamma\cdot\pmb q-\Delta\right)\psi_{\pmb q}=0,\hspace{0.4cm}q\ll 1/a,
  \end{equation}
  which determines $\varepsilon=\varepsilon_{\pmb q}$ ($\varepsilon_{\pmb q}=\pm\sqrt{\tilde c^{\,2}q^2+\Delta^2}$\,). The anisotropy of the Dirac cone \cite{LIU} just 
  rescales momenta in (\ref{46}). The Dirac point is supposed for simplicity to be at $\pmb q=0$. 
  
  The initial Hamiltonian can be considered in terms of tight binding approximation for site functions \cite{MAT}
  \begin{equation}
  \label{47}
  \psi(\pmb r_n)=\int_{BZ}\frac{d^3q}{(2\pi)^3}\psi_{\pmb q}\exp(i\pmb q\cdot\pmb r_n).
  \end{equation}
  The inverse transformation has the form
  \begin{equation}
  \label{47a}
  \psi_{\pmb q}=v\sum_{\pmb r_n}\psi(\pmb r_n)\exp(-i\pmb q\cdot\pmb r_n),
  \end{equation}
  where $v$ is the unit cell volume. In that approximation the neighboring sites $\pmb r_n$ and $\pmb r_n+\pmb\delta_{\alpha}$ interact. Here $\pmb\delta_{\alpha}$ are 
  positions within the unit cell. For graphene $\alpha=1,2,3$ \cite{MAT}.
  
  We denote $\pmb r_n=(\pmb\rho_n,z_n)$. At $\rho_n,z_n\gg a$ one can use the continuous approximation in distance 
  \begin{equation}
  \label{48}
  \left(\gamma^0\varepsilon+i\tilde c\,\pmb\gamma\frac{\partial}{\partial\pmb\rho}+i\tilde c\gamma_z\frac{\partial}{\partial z}-\Delta\right)\psi(\pmb r) =0.
  \end{equation}
  Here the spatial scale $\tilde c/(\varepsilon-\Delta)$ should be larger than $a\sim\tilde c/\varepsilon_0$, where $\varepsilon_0$ is a typical energy in the
  Brillouin zone. This condition, $(\varepsilon-\Delta)<\varepsilon_0$, is fulfilled in a vicinity of the Dirac point. 
  
  Suppose now that only $z_n\gg a$ is large. Then the derivative $\partial/\partial\pmb\rho$ goes over into a difference term containing 
  $\psi(\pmb\rho_n+\pmb\delta'_{\alpha},z)$. Here $\pmb\delta'_{\alpha}$ means in-plane component. Let us apply the transformation (\ref{3}), with $v\ll\tilde c$, 
  to the functions $\psi(\pmb\rho_n+\pmb\delta'_{\alpha},z)$. Then the equation for the spinor $\Phi'(\pmb\rho_n,z)$ acquires the form 
  \begin{eqnarray}
  \nonumber
  &&(\varepsilon-\Delta)\Phi'(\pmb\rho_n,z)+iv\frac{\partial}{\partial z}\sum_{\alpha}c_{\alpha}\Phi'(\pmb\rho_n+\pmb\delta'_{\alpha},z)\\
  &&=-i\tilde c\sum_{\alpha}\left(p_{\alpha}+q_{\alpha}\frac{\partial}{\partial z}\right)\Theta'(\pmb\rho_n+\pmb\delta'_{\alpha},z)
  \label{480}
  \end{eqnarray}
  analogous to Sec.~\ref{tran}. The continuity condition on $z$ is valid at $L=v/(\varepsilon-\Delta)>a$ that is at $(\varepsilon-\Delta)/\varepsilon_0<v/\tilde c<1$.
  In Eq.~(\ref{480}) $p_{\alpha}$ and  $q_{\alpha}$ are the certain $2\times 2$  matrices and the index $\alpha$ runs a few numbers related to nearest neighbors. The 
  right-hand side of (\ref{480}) is the difference analogue of $\pmb\sigma\cdot\nabla\Theta'(\pmb\rho,z)$ in Sec.~\ref{tran}. At $r_c>a$ the scheme (\ref{480}) is reduced
  to the continuous limit (\ref{48}). 
  
  The solution of (\ref{480}) is analogous to (\ref{9a}) with the integration limits $\eta_{1,2}(\pmb\rho_n)$. When these limits are chosen in the singular form 
  like (\ref{10da}), the wave function $\psi(\pmb\rho_n,z_n)$ is singular at the complex values of $\pmb\rho_n$ according to Sec.~\ref{anst}. This justifies the 
  limit of large $z$ for making the integration in (\ref{9a}). The wave function, formed this way, is anomalous. 
  
  The summation on discrete $\rho_{xn}$ (or $\rho_{yn}$) in (\ref{47a}) can be done, as for the Redge summation \cite{LANDAU1}, by the integration along a contour in the 
  complex plane of $\rho_{xn}$. This results in parts of $\psi_{\pmb q}$ depending on chosen singularity positions in that plane. That is $\psi_{\pmb q}$ contains 
  contributions, which cannot be generated by Eq.~(\ref{46}) since in this equation there is no information about the singularities. This means that the expansion 
  (\ref{47}) of the anomalous wave function, with the energy $\varepsilon$, contains harmonics outside the energy surface, $\varepsilon\neq\varepsilon_{\pmb q}$. 
  
  Thus the anomalous $\psi(\pmb r_n)$ is not a superposition of plane waves, which are eigenstates of (\ref{46}) or its extension to entire BZ, according to 
  general properties of anomalous wave functions (Sec.~\ref{anst}). 
  
  \begin{figure}
  \includegraphics[width=4cm]{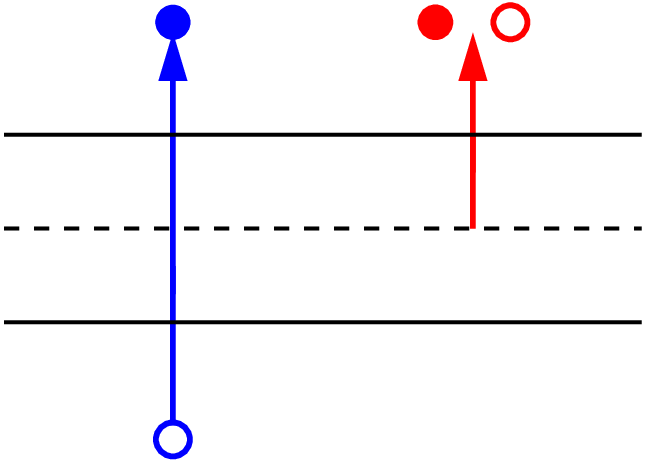}
  \caption{\label{fig1} Creation of the electron-hole pair (left) leaves a vacancy in the valence band. Creation of the electron-antielectron pair (right) occurs from 
  the vacuum.}
  \end{figure}
  
  It follows that two different classes of particles exist in the Dirac insulator. First, this is conventional electrons of conduction and valence bands. Second, 
  this is anomalous quasiparticles, which are electrons and antielectrons. 
  
  The conclusions, drawn for anomalous electrons and positrons in Sec.~\ref{anst}, are valid for anomalous quasiparticles in Dirac insulators. The antielectron, associated with 
  the anomalous electron, appears like the anomalous positron in the Dirac equation. The antielectron is a positively charged quasiparticle, which is not a vacancy in the valence 
  band as shown in Fig.~\ref{fig1}. The positron is also not a vacancy in ``Dirac sea'' but can be created from the electron-positron vacuum. 
  
  We see that the known transition from conventional insulator to topological one \cite{XUI} is assisted by formation of the subsystem of anomalous quasiparticles.
  The pair of anomalous electron and antielectron can be created from the anomalous vacuum keeping the valence band completely filled.
  
  Despite the delicate origin of anomalous particles they are expected to be thermally distributed above the gap in Fig.~\ref{fig1}.
  
  \subsection{Experiments}
  \label{exp}
  By angle-resolved photoemission spectroscopy (ARPES) \cite{XUI} it is impossible to directly identify anomalous particles since they do not exist outside the Dirac
  material. Instead one can use electric current.
  
  The electric conductivity of the Dirac insulator is enlarged due to a contribution of anomalous carriers. Under the electric current $I$ toward the border between 
  the Dirac and a usual insulators anomalous quasiparticles cannot cross this border because of the confinement inside the Dirac insulator. The total current conserves 
  due to a reduction of the electric field inside the Dirac insulator. The jump of the electric field is provided by a surface charge on the border. 
  
  Thus, the macroscopic flux of anomalous quasiparticles should annihilate in a vicinity of the border since a macroscopic backflow, counter to the current, is 
  impossible. This produces the radiation of a macroscopic energy (including electromagnetic one) from the border with the rate $dE/dt\sim I\Delta/e$. We suppose the 
  gap $\Delta$ is roughly on the order of temperature. That electromagnetic radiation recalls the radiative recombination in semiconductors.   
  
  On the other hand, the total energy dissipation is $IV=RI^2+dE/dt$, where $V$ is the voltage applied to the Dirac insulator and the first term is the usual Joule 
  heating. The $I-V$ curve $I=(V-V_c)/R$ follows from hear. That is the electric current appears solely above the threshold voltage $V_c\sim\Delta/e\sim (10-100)mV$. 
  When $V$ is less then the threshold $V_c$, the outer electric field is compensated inside by the surface charge. 
  
  Therefore, in experiments with current across the border between topological and usual insulators (i) the threshold current-voltage characteristics $I=(V-V_c)/R$ is 
  expected, (ii) in the ohmic dependence $I=V/R_0$, before the transition to the topological state \cite{XUI}, the resistivity $R_0$ is bigger than $R$, (iii) one can 
  register the current-induced electromagnetic radiation from the border. 
  
  \section{THE STORY IN SHORT}
  \label{disc}
  A solution of the Dirac equation can be expanded in eigenfunctions characterized by certain quantum numbers. It happens that this conventional class of wave functions 
  is not unique. 
  
  The Dirac equation has a different class of solutions moderated by the generating field $\pmb\eta=\{\eta_1(x,y),\eta_2(x,y)\}$, which is singular at complex values of 
  coordinates. It results in the anomalous wave function roughly of the type $1/(r^2+r^{2}_{c})$. On short distance it is the series of powers of $r/r_c$ corresponding 
  to an expansion in spherical eigenfunctions at small argument. Bur the expansion in this form does not extend to the region $r>r_c$ going over into the series of $r_c/r$ 
  powers. Thus the anomalous wave function is not a superposition of conventional eigenfunctions in the entire space corresponding to a different class of particles referred 
  to as anomalous. As follows, there is the link {\it singularity} $\rightarrow$ {\it particle}. 
  
  The different mathematical aspect is that some stationary solutions of the Hamiltonian wave equation cannot be expanded in conventional eigenfunctions.
  
  The generating field $\pmb\eta$ just maps on the particular shape of the wave function. The anomalous particle is not formed by that field contrary to the electron formed 
  by the Higgs field \cite{HIG}.
  
  The generating field can have various singularities at various complex points. Each $\pmb\eta(x,y)$ produces a particular shape of anomalous state. So that it is not clear
  what are ``quantum numbers'' and ``density of states''. It is also not clear how to define a class of generating fields corresponding to the certain class of anomalous 
  particles. How to define a class? How quantum fluctuations of the generating field influence the anomalous particle? 
  
  Despite the lack of the complete understanding, the certain conclusion can be drawn. The conventional electron-positron pair are created from the vacuum. Analogously the 
  anomalous pair can be created from the anomalous vacuum. 
  
  The topological aspect of Dirac insulators in condensed matter is not a whole story. The field of Dirac insulators, when electrons obey the Dirac like wave equation, has 
  additional features compared to the previous standpoint. 
  
  It turns out that the Dirac insulator is associated with two classes of particles. First, this is conventional electrons and holes with momentum $\pmb q$ and the 
  quasiparticle energy $\varepsilon=\varepsilon_{\pmb q}$. Second, this is anomalous electrons and antielectrons. Their wave functions, of the energy $\varepsilon$, contain 
  harmonics $\pmb q$ outside the energy surface, $\varepsilon\neq\varepsilon_{\pmb q}$. Thus the anomalous state is not a superposition of conventional ones constituting 
  independent particles. 
  
  The antielectton is not a vacancy in the valence band like the positron is not a vacancy in  ``Dirac sea''. The positron can be created from the electron-positron vacuum 
  by a quantum. Analogous electron-antielectron vacuum associates with the Dirac insulator. One can produce there the electron-antielectron pair keeping the valence band 
  completely filled. 
  
  Anomalous quasiparticles in the bulk Dirac insulator can be identified in experiments with electric current across a border between that insulator and conventional one. 
  
  The anomalous subsystem can be also formed in Dirac materials described by the two-component wave function. This is the case of graphen and $d$-wave superconductors. 
  The anomalous formalism is applicable to the massless magnetic monopole mediated by the chiral field \cite{LOSH}.
  
  \section{CONCLUSIONS}
  \label{conc}
  \begin{itemize}
  \item The revealed anomalous electron and positron correspond to the link {\it singularity} $\rightarrow$ {\it particle}. The anomalous particles are of the different 
  class independent of conventional electron and positron.
  \item The different mathematical aspect is that some stationary solutions of the Hamiltonian wave equation cannot be expanded in conventional eigenfunctions.
  \item The Dirac insulator, besides topological aspects, is associated with the anomalous subsystem, where positively charged particles are not holes. Anomalous quasiparticles 
  in the bulk Dirac insulator can be identified in experiments with electric current. 
  \end{itemize}

  \acknowledgments

  I am grateful to J. Engelfried for discussions of related topics. This work was supported by CONACYT through grant 237439. 
  
  \appendix
  
  \section{The Coulomb field}
  Suppose the electron to be acted by the nucleus Coulomb field, which is $U(\pmb r)=-Ze^2/r$ on large distance. When the nucleus charge density is homogeneously
  distributed within the sphere of the radius $r_N\sim 10^{-13}cm$ \cite{BARE}, $U(\pmb r)\simeq -(1-r^2/3r^{2}_{N})Ze^2/3r_N$ at $r\ll r_N$. 
  
  As in the case of free electron, the $\pmb\eta$-generated state is also formed in the potential $U(\pmb r)$. At $r\rightarrow 0$ the results of Sec.~\ref{short} are valid with the
  renormalization $\varepsilon_q\rightarrow\varepsilon_q-U(0)$. The energy $|U(0)|$ is in the $MeV$ range. In the Coulomb field the state also can be marked by $q$.
  
  In the exponent in (\ref{9a}) at $r\gg r_N$ one should change $\varepsilon_q$ by $\varepsilon_q+Ze^2/\sqrt{\rho^2+z^2}$ resulting in
  \begin{eqnarray}
  \nonumber
  &&\exp\left(i\frac{z-\xi}{L}\right)\rightarrow\\
  &&\exp\left(i\frac{z-\xi}{L}+\frac{iZe^2}{\hbar v}\ln\frac{z+\sqrt{\rho^2+z^2}}{\xi+\sqrt{\rho^2+\xi^2}}\right).
  \label{26bc}
  \end{eqnarray}
  At $v\ll c$ the parameter $Ze^2/\hbar v$ is large. 
  
  \section{NUCLEUS ACCELERATION}
  \label{crea}
  A nucleus acceleration leads to the $\pmb\eta$-generated state in the non-inertial frame. The interaction of two ions, separated by the distance $\xi$, can be 
  approximated by the potential
  \begin{equation}
  \label{346d}
  V(\xi)=E_0\left(\frac{a}{\xi}\right)^{12}-\frac{e^2}{\xi},
  \end{equation}
   where the first part is the Lennard-Jones repulsion with $a\sim 10^{-8}cm$ and $E_0\sim 1\,eV$. This potential is almost
  \begin{equation}
  \label{346f}
  V(\xi)=E_0\left[\left(\frac{a}{\xi}\right)^{12}-\frac{a}{\xi}\right].
  \end{equation}
  
  The function $\xi(t)$ is determined by the equation $M{\dot\xi}^2/2+V(\xi)=E$, where $E$ is the ion energy and $M\sim 10^4m$ is the ion mass. When $E\ll E_0$, 
  $\dot\xi\sim\sqrt{E_0/M}\sim 10^{5}cm/s$. At $E_0\ll E$ the velocity  $\dot\xi\sim\sqrt{E/M}$.
  M
  Suppose the electron to be acted by the electrostatic field $U[\pmb r-\pmb\xi(t)]+U_1[\pmb r-\pmb\xi_1(t)]$ of two colliding ions localized at time variable positions 
  $\pmb{\xi}(t)$ and $\pmb{\xi}_1(t)$. In the region close to $\pmb r=\pmb\xi(t)$ the potential $U_1$ plays a minor role since it is localized apart. For this reason, we 
  consider $U_1$ as a perturbation accounting it at the final step.  
  
  The Dirac equation has the form now
  \begin{equation}
  \label{341a}
  \left[\gamma^0\left(i\frac{\partial}{\partial t}-U[\pmb r-\pmb\xi(t)]\right)+ic\pmb\gamma\cdot\nabla -mc^2\right]\psi(\pmb r,t)=0.
  \end{equation}
  The nucleus displacement $\pmb{\xi}(t)$ is a slow varying function compared to the fast electron motion. At every moment $t_1$ one can approach 
  $\pmb{\xi}(t)\simeq\pmb{\xi}(t_1)+\dot{\pmb\xi}(t_1)(t-t_1)$. 
  
  Let us consider the solution $\psi(\pmb r,t)$ of (\ref{341a}) at the moment $t_1$, from the standpoint of the frame moving with the velocity $\dot{\xi}(t_1)$ directed along 
  the $z$-axis. The transition to the non-inertial frame, where the the potential is static, modifies the Dirac equation \cite{FEL}. In the limit $\dot{\xi}\ll c$, with the 
  adiabatically varying $\dot{\xi}$, one can just make the change of variables $\pmb R=\{x,y,z-\dot\xi(t_1)t\}$ and $t'=t-\dot\xi(t_1)z/c^2$. With the substitution 
  $\psi[\pmb r(\pmb R,t'),t(\pmb R,t')]=\exp(-i\varepsilon t')\psi(\pmb R)$ the Dirac equation (\ref{341a}) takes the form
  \begin{eqnarray}
  \nonumber
  &&\Bigg[\left(\gamma^0-\frac{\dot\xi}{c}\gamma_z\right)\varepsilon-mc^2-\gamma^0U(R)\\
  &&+ic\left(\pmb\gamma\cdot\nabla-\frac{\dot\xi}{c}\gamma^0\frac{\partial}{\partial z}\right)\Bigg]\psi(\pmb R)=0.
  \label{5a}
  \end{eqnarray}
  We suppose $U(R)$ to be isotropic. 
  
  In the approach used $\dot\xi$ adiabatically depends on time. One can easily show that non-adiabatic effects lead to the modification of the potential 
  $U(R)\rightarrow U(R)+m\ddot\xi z$. As follows from (\ref{346f}), $m\ddot\xi=-V'(\xi)m/M$. The maximal value of $V'(\xi)$ is $E_0/a$, which is of the atomic scale. 
  Since $m/M\sim 10^{-4}$ and $V'(\xi)$ is non-zero at $t\lesssim 10^{-13}s$, non-adiabatic effects in (\ref{5a}) are negligible. Eq.~(\ref{5a}) corresponds to a lag
  of the electron behind the moving nucleus. 
  
  Eq.~(\ref{5a}) is the same as (\ref{5}) with $-\dot\xi(t)$ instead of $v$. Thus, $\dot\xi(t)$ determines states adiabatically varying in time. Anomalous states are
  among them.
  
  This situation with the controlled velocity parameter takes place in experiments with strongly accelerated nuclei. It can be, for example, in a high voltage discharge 
  in gases \cite{OGI1,OGI2} or liquids \cite{URU,PRI}. It seems that these experiments require an attention since for \cite{OGI1,OGI2} ``known fundamental interations
  cannot allow prescribing the observed events to neutrons'' \cite{BAB}. The strong nucleus acceleration occurs also in the natural lightning, where high energy processes
  are revealed \cite{ENO}.
  
  There is also the wave function with no electron lag in the frame, where the nucleus is at rest. This wave function follows the additional transformation (\ref{3}) 
  with $v(t)=\dot\xi(t)$. The resulting wave equation differs from the usual Dirac form by the small non-adiabatic term $\ddot\xi/c$.

  \end{document}